\documentclass{aastex}

\shorttitle {Supernebulae in II~Zw~40}
\shortauthors{Beck et al.}

\begin{document}

\title{Radio-Infrared Supernebulae in II~Zw~40}

 \author{Sara C. Beck\altaffilmark{1}, Jean L.Turner\altaffilmark{2}, 
Laura E. Langland-Shula\altaffilmark{3},
David S. Meier\altaffilmark{2,4}, 
Lucian P. Crosthwaite\altaffilmark{2,5},  and
Varoujan Gorjian\altaffilmark{6}}

\altaffiltext{1}{Department of Physics and Astronomy, Tel Aviv University, 
Ramat Aviv, Israel email: sara@wise1.tau.ac.il}
\altaffiltext{2}{Department of Physics and Astronomy, UCLA, Los Angeles,
CA 90095-1562 email: turner@astro.ucla.edu}
\altaffiltext{3}{Department of Astronomy and Astrophysics, University
of California, Santa Cruz, CA 95064 email: laura@ucolick.org}
\altaffiltext{4}{present address: Department of Astronomy, University
of Illinois, Urbana, IL 61801 email: meierd@astro.uiuc.edu}
\altaffiltext{5}{Astute Networks, 16868 Via Del Campo Ct. Suite 200,
San Diego CA 92127 email: pat@astutenetworks.com}
\altaffiltext{6}{Jet Propulsion Laboratory, California Institute of 
Technology, MS 169-327, Pasadena, California email: vg@jpl.nasa.gov}

\begin{abstract}
We report subarcsecond-resolution VLA and Keck mid-infrared imaging of the dwarf
starburst galaxy II~Zw~40. II~Zw~40 contains a bright,
compact thermal radio and infrared source with all the the characteristics
of a collection of dense HII regions ionized by at least
14,000 O stars.  The supernebula is revealed to be multiple sources
within an envelope of weaker emission. The radio
emission is dominated by free-free emission at 2~cm, and the 
spectrum of this emission appears to be rising.  This suggests that
 the free-free emission is optically thick at 2 cm, and that the 
individual HII regions
are $\sim$1 pc in size. This complex of ``supernebulae"
dominates the total infrared luminosity of II~Zw~40, although the
radio source is less than
$\sim$150~pc in diameter.   Multiple super star clusters appear
to be forming here,  the much larger analogues of large
Galactic HII region complexes.

 \end{abstract}


\section{Introduction: Infrared and Radio Supernebulae in Starbursts}
Starburst galaxies are distinguished by the concentration, and not simply
the quantity, of star formation. Optical and UV studies with the
resolution of the HST have found star formation regions to contain
super-star clusters (SSCs), as populous and as dense as globular clusters but
with ages of only a few million years. Observations in the radio and 
infrared are finding the even younger and obscured counterparts of these star 
clusters. Traced by optically thick free-free emission 
at centimeter wavelengths, radio/infrared ``supernebulae" 
\citep*{THB98,TBH00,G01} or ``ultra-dense HII regions"
\citep{KJ99} have been discovered in nearby starburst galaxies. The
discovery of these regions 
has been due to the development of subarcsecond imaging 
in the infrared and radio. 
These HII regions are bright and compact, excited by immense clusters 
of hot young stars, or ``starforming clumps" \citep{B01}.  
The sources have in common all of the signs of young star formation:
 high gas density \citep[often $\rm >10^4~cm^{-3}:$][]{THB98,KJ99,Tar00,M02}, 
a cm-wave continuum with spectral index from -0.1 to +1.5, substantial
extinctions \citep*[Av $>\sim 10$ mag:][]{K89,H90},
and very strong mid-infrared emission \citep*{G01,B01,Dale01,VJC02}.  
It is not just that these
supernebulae are very luminous in the infrared, they may actually dominate
the infrared emission of the entire galaxy. In the nearby dwarf
galaxy NGC~5253, for example,
at least 75\% of the mid-infrared emission seen by IRAS and  
25\% of the total bolometric luminosity appear to come from one 
obscured star cluster less
than 2~pc in diameter \citep{G01}. In another starbursting dwarf,
He 2-10, essentially all
the mid-infrared emission is from star-forming clumps in a 200 pc by 50
pc disk \citep*{B01,VJC02}. 
If infrared emission traces and quantifies star formation, as
has been thought since IRAS, then since these clusters or supernebula
are {\it the} sources of infrared, they are also {\it the} starburst.

II~Zw~40 is a starburst dwarf galaxy 10.5 ($\rm H_o$/75) Mpc away.  
It appears to be
the result of a collision of two smaller galaxies \citep*{BST,BK88,
Van98}. 
At optical wavelengths
it is dominated by one bright star cluster \citep{S70}, which is also a strong
source of $H\alpha$ \citep{SS70} and associated nebular
lines \citep{WR}, has the Wolf-Rayet feature \citep{VC}, and in
general has all the signs of a young star formation region.  The
infrared spectrum is dominated by high-excitation lines such as [NeIII]
and [SIV] \citep{R91, Thornley,B02}. The radio continuum emission 
is nearly 
all thermal emission \citep{KWB91,KWT84,J78,D93}, and the emission
is relatively compact \citep{WWB,SW,KWB91}. Like the radio continuum,
the H$\alpha$ \citep{SS70} 
and Brackett $\gamma$ lines \citep{WWB,MO,JL,Ve96,DSW}
are bright in this galaxy . Molecular gas is
conspicious by its absence; \cite{S92} remark on its unusually
high star formation efficiency. 
In these respects, II~Zw~40 resembles NGC 5253 
\citep{Cr99,THB98,TBH00}, probably
the youngest, most massive, and most extreme example of an obscured
young super star cluster or supernebula in the local universe. 

Motivated by
the  galaxy's similarity to NGC 5253, we imaged II~Zw~40 at 6, 3.6 and
2 cm using the Very Large Array\footnote{The National Radio Astronomy Observatory is a facility of the National Science Foundation operated under cooperative agreement by Associated Universities, Inc.} 
in its A configuration for the maximum angular resolution, and at
$11.7\mu$m with the Long Wavelength Spectrometer (LWS) at the 
W. M. Keck Observatory.\footnote{The W. M. Keck Observatory is operated
as a scientific partnership among the California Institute of Technology,
the University of California, and the National Aeronautics and Space
Administration. The Observatory was made possible by the generous 
financial support of the W. M. Keck Foundation.}  The observations and
results are described in the next section and the nature of the source
in $\S3$, in which we also compare the supernebula in II~Zw~40 to other
dwarf starbursts and discuss the significance of the structure seen in
the radio maps.

\section{Observations and Results}

\subsection{Radio Images}
II~Zw~40 was observed at 6, 3.6 and 2~cm on 23 August 1999, using the
VLA in the A configuration. The observations began in late morning
with reasonable summer weather. During the run
the 12 GHz site monitor rms phase 
rose from 12 to 20$^\circ$. Absolute fluxes were 
calibrated using 3C286, with fluxes of 3.45 Jy, 5.18 Jy, 
and 7.49 Jy at 2, 3.6, and 6 cm, respectively. We estimate absolute flux 
(scaling) uncertainties of $\sim$ 5\% at 6 and 3.6 cm, and $\sim$7\% at
2~cm due to source structure and variability in 3C286. Missing
antennas in the inner array contributed to these uncertainties. 
For the 2~cm data, we used fast-switching with a 
50s-calibrator/100s-source cycle time. We spent 6 minutes, realtime,
 on source at
6~cm and 3.6~cm, and 79 minutes, realtime, on source at 2~cm. 
However, given the
fast-switching duty-cycle, the actual on-source time at 2~cm
was about 44 minutes. The nearby calibrator 0552+032 was
used for the fast-switching cycle; we measure a flux of 0.53 Jy 
for this source. The fast-switching
cycles were bracketed with standard calibration scans on the stronger
calibrator 0532+075, for which we measure a flux of 1.65 Jy. 
We applied referenced pointing during the run, with pointing scans 
taken at 3.6 cm
every hour on 0532+075. In this way we were able to obtain reasonable
images with the A~configuration at 2~cm in daytime summer observing. 
We estimate, from 
our mapping of the calibrators, that the ``seeing" at 2~cm with our
fast-switching cycle is less than
$<$0.01\arcsec, which is the upper limit we fit to the deconvolved sizes of
both calibrators. This is less than a tenth of a beam. Due to the lack of
short spacings, the images are insensitive to structures that are 
more extended than $\sim$ 30 times the beam. This approximate
limit, $\theta_{max}$,
is listed for each wavelength in Table 1, along with 
 beamsizes and rms noise levels.

Images at each wavelength are in Figure 1. These maps were made
with natural weighting. The emission is clearly
resolved, especially in the highest resolution 2~cm map. The position (J2000)
of the 2~cm peak of 0.64 mJy/beam is 
$\alpha = 05^h~55^m~42$\fs614~$\pm$0\fs 001, 
$\rm \delta = 03^\circ~23^\prime~32$\farcs 03~$\pm$0\farcs 01. 
The main source is
kidney-shaped, hinting that there is further unresolved substructure
on sizescales less than 0\farcs 4 (20~pc).
 There is a secondary 2~cm
peak, with a flux of 0.54~mJy/beam, located 0.23\arcsec\ (11~pc) to the 
east of the main source, at 
$\alpha = 05^h~55^m~42$\fs629~$\pm$0\fs 001, 
$\rm \delta = 03^\circ~23^\prime~32$\farcs 04~$\pm$0\farcs 01. 
The 6~cm and 3.6~cm
sources are consistent with the structure seen at 2~cm.

Peak fluxes and other information for naturally-weighted maps 
are presented in Table~1. Total mapped fluxes were obtained
by integrating the flux radially outward from the peak until it
reached a maximum using AIPS task IRING. These mapped fluxes are
presented in Figure~2 with the single dish fluxes and other
interferometric fluxes from the literature. Most of the
6~cm and 3.6~cm flux is contained within a radius of 5-6\arcsec.
The 2~cm flux is confined to the inner 1\farcs 5 (75~pc) radius region.
These sizescales are similar to $\theta_{max}$, the 
angular size at which undersampling effects set in. 
Sixty to eighty percent of the single dish emission 
is concentrated within this central few arcsecond
portion of II~Zw~40.
 
The spectral index of radio continuum emission, which is defined as
$\alpha$ in $S_{\nu}\propto{\nu}^{\alpha}$, is -0.1 for pure thermal emission,
some larger negative value, typically -0.6 to -1.2 for non-thermal
emission from supernovae events and remnants, and takes on positive
values when the radio emission is optically thick. We cannot formally
derive the spectral index because the naturally-weighted beams at different
wavelengths are not matched: the maps at the shorter wavelengths
will lose relatively more flux due to undersampling. The spectral index
of the overall radio emission in II~Zw~40 has been discussed at length in 
\citet{WWB,KWB91,D93}. These studies indicate that the cm-wave
emission of II~Zw~40, even at 20~cm, is mostly thermal, 
and that thermal emission is 
overwhelmingly dominant at wavelengths less than 6~cm.

\subsection{12$\mu$m Image}
II~Zw~40 was observed on the night of 17 February 2000 using the LWS
\citep{JP93} on
the Keck 1 telescope in imaging mode. The resulting image is 
shown in Figure
1d (greyscale) and in Figure 3. The exposure was 486 seconds 
through a $1\mu$m wide filter centered
at $11.7\mu $m.  $\beta$ Gemini and $\alpha$ Bootes were used for
calibration. The night was humid and non-photometric; scatter in the
standards was 7\% in an hour but as high as 18\% over the night, so we
conservatively assign 15\% uncertainty to the infrared flux.  The pixel
size is $0.08\arcsec$ and the chip $128 \times 128$; the actual resolution,
based on the standard star images, is $0.3\arcsec$ to $0.5\arcsec$.  The
infrared source observed is either not or only slightly resolved; the
FWHM is  $\sim 0.5\arcsec$. At this resolution
 the source cannot be distinguished from
circularly symmetric.  

The total flux at $11.7\mu $m is 0.24 Jy.  This
agrees with the 0.22 Jy \citep{RL72} observed with a $6\arcsec$
aperture and is more than half of the 0.46 Jy seen by IRAS with a beam
that took in the entire galaxy. The bright mid-infrared emission 
indicates that the compact radio source is not a supernova remnant.
{\it The II~Zw~40 source has all the radio and infrared behavior of an HII
region ionized by young OB stars.}  The ratio of mid-infrared to thermal
2 cm radio emission is $\sim 20$ for this compact radio/IR source, 
on the low side of the range
expected for an HII region \citep{H90} but quite out of the range of
supernovae. The value of 12$\mu$m/6~cm is only a factor of two greater 
than the value expected
on the basis of Ly$\alpha$ heating alone \citep{Genzel82}, and it
is nearly an order of magnitude smaller than it is in the otherwise
very similar supernebula in NGC~5253 \citep{G01}. The 
presence of an infrared excess (IRE) over the value expected from
Ly$\alpha$ heating is generally attributed to factors such as 
the competition of dust and gas for UV photons and the absorption
of longer wavelength light by dust. The conditions in
II~Zw~40---low metallicity and a dominant OB population---may
not be conducive to the ``normal" IRE.

We list measured IR fluxes from the literature
for II~Zw~40 in Table~2, including large-aperture IRAS fluxes, 
and we plot the IR SED in Figure 2.
We fit the SED of II~Zw~40 with three dust components, adopting
dust emissivities proportional to $\nu^\beta$, where $\beta=1.5$.
The fits are also illustrated in Figure~2.

Our 11.7$\mu$m flux of 0.24 Jy agrees
well with the fluxes measured by other groups \citep{RL72,WWB} in 
few-arcsecond apertures, and suggests that the mid-IR flux of the central
starburst region 
 is confined to the 0.5\arcsec\ (25 pc) source in
our image. The spectrum of this source is rather flat in the mid-IR.
It therefore requires fairly warm emission; we can fit it
  with a  dust component of temperature
T$\sim 200$K, giving a luminosity of $\rm L\sim 2.4\times10^8~L_\odot$
for this component (see Figure 2).
 Dust emission from the supernebula in NGC~5253 
and from the dwarf galaxy SBS 0335-052 also indicate hot (200K) dust,
based on high resolution
imaging at 11.7$\mu$m  and 18.7$\mu$m \citep{G01,Dale01}.

We can estimate a total $\rm L_{IR}$ for II~Zw~40 by including
IRAS fluxes. The $25\mu$m IRAS flux of II~Zw~40 is due to hot dust with
a temperature of $\sim$100K, we estimate, to explain both the 
$25\mu$m and $60\mu$m components--again, like NGC~5253. This 
component (Figure 2) contributes about
$\rm 7.2\times 10^8~L_\odot$, about three times what the 200K dust
contributes. The dust emission is unusually warm in this galaxy
 \citep{V93}, although the cool (42K) dust
does contribute about half ($\rm 9.4\times 10^8~L_\odot$) of the 
total IR luminosity. The  $100\mu$m
is also complicated by the presence of Galactic cirrus. From the 
dust emission out to a wavelength of  $100\mu$m we obtain
a total infrared luminosity of $\rm L\sim 1.9\times10^9~L_\odot$.

\subsection{Radio Analysis of the Source: Thermal and Nonthermal
Emission and Sizescales}

There is a wealth of radio continuum observations of II~Zw~40, both
single dish and interferometric, and thus much information on the
distribution of thermal and nonthermal emission in the source
\citep{J78,KWT84,SW,KWB91,D93}.\citet{Ve96} and \citet{JL} 
also discuss the radio structure and the thermal/nonthermal
flux ratio. Since our A array images do not detect
large spatial structures,  it is useful to compare them 
to lower resolution maps and single
dish fluxes which are sensitive to extended emission.  

At 6~cm, the lower resolution VLA images of \citet{WWB} and \citet{SW}
revealed 16 mJy of flux within a 20\arcsec\ region (corresponding to their
undersampling radius). The total 6~cm single dish flux is about 21-22 mJy 
\citep{J78,KWB91}. We detect 15$\pm$1~mJy at 6~cm, 68\% of the
total single dish flux. \citet{Ve96} predict, based 
on their extinction-corrected H$\alpha$ flux, a 6~cm flux of 15.7~mJy
for a 15\arcsec\ aperture. \citet{JL} also conclude that their near-IR
and Br$\gamma$ fluxes are in excellent agreement with the VLA fluxes.
From this we conclude that at least 70\% of the single-dish
6~cm emission is thermal, and confined to a region of less than 6\arcsec\ 
(300 pc) in extent.  The remaining 30\% of the single dish 6 cm emission
forms a diffuse extended 
\citep[$>$15\arcsec, or 750 pc, based on the lower 
resolution VLA maps of ][]{WWB} halo of thermal emission 
about this central, bright source. 

We detect a similar percentage of the total single dish emission
at 2~cm. Our flux of 14 mJy is 
$\sim$75\% of the estimated single dish emission of 18-19 mJy at 15~GHz
\citep[extrapolated from the 10.55, 10.7, and 24.5 GHz measurements of ]
[]{J78,KWB91,KWT84}. Our 2~cm flux
is consistent with the VLA fluxes of \citet{WWB} and \citet{SW} of 12 mJy
from lower resolution observations. 
The 14 mJy of 2~cm flux is confined to within a 
3\arcsec\ diameter region; of this, 10 mJy is found in the
inner,  1.5\arcsec\ diameter,
bright ``core" region visible in the contour plot of
Figure~1. The remaining 4-5 mJy of single dish flux
that is not detected in any of the VLA maps must either be very
diffuse, extended over sizescales $>$20\arcsec\ (1 kpc), or consist
of a spatially extended collection of low flux  sources.

The total 2~cm flux
is much greater than the sum of the two peaks seen in the 
naturally-weighted maps. For our flux of
10 mJy for the inner 1.5\arcsec\ (75 pc) core region, the required Lyman
continuum fluxes are at least
$\rm N_{Lyc}=1.0\times10^{53}~s^{-1}$, and higher
if the emission has significant optical depth that it is partly thick at 
2 cm. The total mapped 2 cm flux of 14~mJy, which is contained
within the inner 3\arcsec,  
requires $\rm N_{Lyc}=1.4\times10^{53}~s^{-1}.$

\subsection{``Matched-Beam" Radio Images}

We cannot directly compare fluxes at different wavelengths
for the naturally-weighted maps of the previous section. This is because
the missing short baselines scale to different angular resolutions
at the different wavelengths: the spatial scale at which emission
is resolved out in the 2~cm maps ($\rm \theta_{max}\sim 3-4$\arcsec) is
three times smaller than for the 6~cm map ($\rm \theta_{max}\sim 10$\arcsec).
So the maps at each wavelength are missing a different fraction of the 
true total flux. 

It is possible, however, to make maps at different wavelengths for which 
fluxes can be directly compared. For this we must restrict the baselines
in the inner (u,v) plane in the 6 cm and 3.6~cm maps to match
the 2~cm (u,v) restriction. With the current data set, we eliminated all baselines
for which B/$\lambda<20,000$. The resulting 6~cm and 
3.6~cm images are noiser than before, but their sensitivity to extended
structure is close to that of the 2~cm image.  We then convolved these 
``nearly identical" 
beams to the Gaussian beamsize of the 6~cm map, 0\farcs 88 $\times$
0\farcs 43, p.a. 52$^\circ$, and compared them. 

Total mapped fluxes for these (u,v)-restricted maps, which resemble
the naturally-weighted 6~cm map in appearance, are presented in the last 
column of Table~1. At 6 and 3.6 cm 
the ``matched beam" images 
contain less flux than do the naturally -weighted images:
as the undersampling sizescale, $\theta_{max}$, decreases, more 
of the flux is resolved out.
Of the 15 mJy of flux in the naturally-weighted 6~cm map 
($\theta_{max}\sim 10$\arcsec), only 9 $\pm$1.5 mJy remains in the ``matched
beam" image; of the 12 mJy of flux in the naturally-weighted 3.6~cm
image, only 10$\pm$1.5 mJy is left in the ``matched beam" map.   
But since the U
band map is already restricted with respect to short spacings,
the ``matched beam" flux is the same as for the naturally-weighted
map; the 2 cm flux is 14 mJy. 

Comparison of the ``matched beam'' fluxes indicates that
the spectral index of the compact ($\theta <4$\arcsec ) emission in II~Zw~40 
is rising from 6~cm to 2~cm, with a value between +0.4 and +2. 
This strongly suggests that a significant
fraction of the 2 cm free-free emission is not only compact, but optically
thick.

\subsection{Highest Resolution Image: The Compactest Structure}

To further investigate the compact structure we have  
mapped the 2~cm emission with the highest possible resolution, 
and the result is presented
in Figure~4. This map was made with uniform weighting, which places greater weight on the
longer baselines. For standard VLA configurations, this sharpens the
beam by roughly 50\%. The noise in this image, 0.17 mJy~$\rm beam^{-1}$,
is a factor of two higher than in the naturally-weighted map,
which is why less emission is seen. The uniform 
beam is 0.14\arcsec $\times$ 0.12\arcsec, p.a. 44$^\circ$.  

The uniform image
reveals 3-4 peaks within the central emission region, all of
which are 0.59-0.60 mJy/beam, 3.5$\sigma$. The three strongest
peaks are
separated by 0\fs015, or 11 pc. They are at (J2000)
$\alpha =5^h 55^m$ 42\fs630, 42\fs615, 42\fs601~$\pm$~0\fs 001,
$\delta$=03$^\circ$ 23\arcmin\ 32\farcs01, 32\farcs02, 
32\farcs02~$\pm$~0\farcs 01, from east to west, respectively. 
The first two sources listed also appear 
as the secondary and main peak in the lower-resolution 2 cm image of Figure~1. 
The fourth source is 6 pc to the north of the westernmost source of
 the triplet, also 6 mJy/bm, at 
$\alpha $=5$\rm ^h$ 55$\rm ^m$ 42\fs601~$\pm$~0\fs 001,  
$\delta$=03$^\circ$ 23\arcmin\ 32\farcs 13~$\pm$~0\farcs 01. 

These sources may simply be the
peaks of the more extended distribution, although there is
an increase in brightness temperature with decreasing beam. The
2~cm naturally-weighted map has $\rm T_b\sim 120$K, while the
uniform map, with a beam a factor of 1.7 smaller in area,
 has $\rm T_b\sim 200$K. This suggests that
there is indeed some unresolved structure. The brightest peak in the
uniform map, at $\alpha =5^h 55^m$ 42\fs 525,  $\delta$=
03$^\circ$ 23\arcmin\ 32\farcs 88 has a peak flux of 0.76 mJy/beam,
$\sim 5\sigma$. This peak is located about 1.5\arcsec\ to the
northwest of the center of the triplet. There is little extended
emission associated with this source, as it is not strong in
the lower resolution images.

We can estimate the source sizes using the spectrum 
and the intensities observed here.
Peak flux densities of $\sim$ 0.6 mJy/beam for
the quadruplet correspond to brightness temperatures of $\sim$200~K for
the 0\farcs14 by 0\farcs12 beam. In order for this brightness to 
be consistent with the rising spectrum, and optically thick
emission, the source must be confined to a
region smaller than the beamsize. For an
electron temperature of 13000K, as determined from optical 
lines by \citet{WR}, and
for $\tau\sim 1$,  we estimate that a source diameter of 
d$\sim 1$~pc would produce the observed brightness. 
The emission measure implied by a turnover frequency of 15~GHz
is about $\rm 10^9~cm^{-6}\,pc$, and the corresponding nebular
density would then be $\rm \sim 3-4\times 10^4~cm^{-3}$. Emission measures
and densities of these magnitude are seen only in the youngest
HII regions in the Galaxy. This  suggests
that these nebulae are very young. The dynamical time for
the expansion of an HII region to this size at a sound speed
of 10 km/s is $\sim 10^5$ yrs; its size is another suggestion of its likely
youth.

For each source within the quadruplet, 
we find a lower limit to the Lyman continuum 
of  $\rm N_{Lyc}\sim6\times10^{51}~s^{-1}$ for $\tau=0$ (optically
thin),  
$\rm N_{Lyc}=0.8-1\times10^{52}~s^{-1}$ for $\tau=1$,
and higher if $\tau >1$. 

\subsection{Comparison with Brackett $\gamma$; Extinction in the Nebula}

Brackett $\gamma$ fluxes have been obtained and analyzed by several
groups, including \citet{WWB} and \citet{JL}. \citet{Ve96} and
\citet*{DSW} have
actually imaged Brackett~$\gamma$. We can
compare our high resolution free-free fluxes with their results to 
estimate extinctions.

We measure 14 mJy of free-free emission at 2~cm in II~Zw~40, from a
region $\sim$3\arcsec\ in diameter. Using the recombination emissivities 
of \citep{HS87} for 12500K, we would 
predict on the basis of the 2~cm free-free flux a Br~$\gamma$ flux
of $\rm S_{Br\,\gamma}^{pred}=1.2\times 10^{-13}~erg~s^{-1}~cm^{-2}$. 
This can be compared to the 
$\rm S_{Br\,\gamma}^{obs}=0.3-0.4\times 10^{-13}~erg~s^{-1}~cm^{-2}$ 
measured by
\citet{Ve96} in an aperture of 3-4\arcsec, and 
$\rm S_{Br\,\gamma}^{obs}=0.4\times 10^{-13}~erg~s^{-1}~cm^{-2}$ 
measured by
\citet{DSW}.  The extinction at 2.2~$\mu$m
implied by the difference is $\rm A_K\sim 1-1.2$, for a visual 
extinction of $\rm A_V\sim 8-10$ mag \citep[adopting the extinction curve
of][]{RL85}. Based on comparisons of H$\alpha$ and Br$\gamma$ 
fluxes, \citet{JL}, \citet{Ve96}, and \citet{DSW}
 estimate $\rm A_K\sim 0.32-0.36$,
a factor of 3 lower than we find. It is likely that the apparent
contradiction is caused by the fact that much of the extinction is high 
and internal
to the nebula. Infrared line ratios frequently give higher extinctions
than do visible line ratios \citep{B86}. $\rm A_K\sim$ 1 mag is seen in many
starbursts of this luminosity \citep{H90,K89}.

We interpolate a single dish flux of 18.5 mJy at 2~cm from the 
2.88~cm and 1.18~cm measurements of \citet{KWB91} and \citet{KWT84}.
For apertures larger than $\sim$4\arcsec\ we would predict
a  Br$\gamma$ flux of $\rm 
S_{Br\,\gamma}^{pred}=1.6\times 10^{-13}~erg~s^{-1}~cm^{-2}$,
which can be compared with the observed flux of 
$\rm S_{Br\,\gamma}^{obs}=7\times 10^{-14}~erg~s^{-1}~cm^{-2}$ measured by \citet{Ve96}
for a 15\arcsec\ aperture and \citet{MO} in a 6\arcsec\ x 6\arcsec\
aperture. From these fluxes we estimate $\rm A_K\sim0.96$, or
$\rm A_V\sim 8$ for emission on larger ($\rm r > 200~ pc$) sizescales.

\section{The Nature and Stellar Population of the Source}

The infrared and radio emission from II~Zw~40 has the form of a
concentrated 1.5\arcsec\ (75 pc) diameter core containing at least 
half (10 mJy) of the large aperture flux, within a slightly larger (3\arcsec, 
150 pc diameter, 14 mJy)
region of lower-level (2-2.5$\sigma$) extended emission,
and diffuse thermal emission extended on larger scales which cannot
be detected in these images. 
Our further discussion will concentrate on the 
compact thermal sources detected in these maps.

 The number of stars needed to ionize the radio
source can be estimated from the total ionization of $\rm 1.4\times 10^{53}~sec^{-1}$. This requires $\rm \sim 1.4\times 10^4$
typical O7 stars over the central 3\arcsec\ (150 pc)  region,
of which $10^4$ are confined to the inner 75 pc core. 
If the stars follow the nebular emission---as seems to be
the case in NGC~5253---then from its appearance in the highest resolution map
 II~Zw~40 is currently forming at least four pc-sized
clusters with $\sim$ 600 O stars each. For a Salpeter ZAMS IMF down
to $\rm 1~M_\odot$,
these four clusters would have masses of $\sim 10^5~M_\odot$,
and higher if the IMF extends to below $\rm 1~M_\odot$. The
estimated mass for the entire 150 pc region would be at least
$\sim 2\times10^6~M_\odot$ (O3 to G).

The luminosity of this large population of young stars is 
consistent with the IR spectrum, in which the mid-infrared
contribution is substantial. 
The total observed IR luminosity out to 100$\mu$m
 is $\rm L_{IR}\sim 1.9\times 10^9~L_\odot$. We estimate,
using a ZAMS Salpeter IMF with upper mass cutoff of O3 
\citep[following][]{THB98}
that the ratio of ZAMS luminosity/Lyman continuum photons
is $\rm N_{Lyc}\sim 2\times 10^{-44}~s^{-1}$. Then our Lyman continuum rate 
of $1.4\times 10^{53}~s^{-1}$ implies an OB luminosity of 
$\rm L_{OB}\sim 3\times 10^9~L_\odot$. 
This OB luminosity is very similar to $\rm L_{OB}$ obtained by 
\citet{Ve96}
on the basis of emission line excitation. $\rm L_{OB}$ may
exceed $\rm L_{IR}$, although there are uncertainties in the
SED and in the fit, and in 
$\rm L_{OB}$. What is clear, however, is that the current starburst
is responsible for the infrared luminosity; even  a ZAMS OB 
luminosity appears to overpredict $\rm L_{IR}$. 
This suggests that the starburst is exceedingly young, or that there
are unusually massive stars, O2 or above, as might be suggested
by comparison with the smaller and less extreme
 cluster R136 \citep{MH98}. The
thermal radio and infrared source, which we call the supernebula because
of its similarity in all but size to a normal HII region, provides all
of the infrared luminosity in II~Zw~40. In fact, in this galaxy, the
infrared luminosity may fall short of measuring the true star formation
rate.

We would like to emphasize a point made by 
\citet{Ve96}: the radio to
IR correlation seen in many galaxies does not hold in II~Zw~40, since the
cm-wave fluxes fall short of the value predicted by the
radio/IR correlation by factors of 6-10. Given that the IR excess in
this galaxy appears to be low, based on the thermal radio fluxes,
this discrepancy is even more notable. There is just not much nonthermal
synchrotron emission in this BCD for the magnitude of its star
formation. This is consistent with the starburst being extremely young.
However, the very similar starburst in NGC~5253 shows a similar lack 
of nonthermal emission
even though it has clusters estimated to range from 2.5-50~Myr
\citep{C97,Tre01}. It is
unclear if the lack of nonthermal synchrotron emission from these
starbursting dwarfs is due to the youth of the starburst or to
some other cause, such as the lack of a global magnetic field
\citep{THB98}.

\subsection{The Structure of the Starburst in II~Zw~40:
Multiple Star Clusters?}

The starburst in II~Zw~40 clearly has a complex radio
 structure, best seen in our 2 cm map (Fig. 1a and Fig. 3).  In the 
highest resolution uniform maps
there are at least four sources visible, within a region
of emission that extends to a diameter of 150 pc.  
Under the very rough assumptions
of the earlier sections we estimate that the starburst 
contains 14,000 O7 star equivalents. The starburst appears to be  
made up of four compact, 1-pc diameter nebulae 
excited by clusters each containing $\sim$600 O stars and
$\rm \sim 10^5~M_\odot$. The individual nebulae/clusters 
are separated by 10-12 pc.  
These characteristics are similar
to the supernebulae in NGC~5253, He 2-10, and
NGC~2146 \citep{B96,THB98,KJ99,TBH00,
Tar00,B01}. 
But it should be noted that what we refer to as
``clusters" and ``sources" are not isolated and well-defined emission 
regions, like globular 
clusters in our Galaxy. Rather, these are emission peaks within an extended
region of emission.  In this sense the nebulae resemble large 
Galactic HII region complexes such as W49 and W51.  
It would be very worthwhile to re-observe this galaxy with the
highest possible resolution and better S/N. 

If the O stars deduced from the radio flux formed alongside the usual
number of smaller stars, there will be a total mass of 
$\sim 2\times10^6 M_\odot$ in young
stars within the 150 pc volume of the starburst, and an estimated
$10^6-10^7$ young stars. This stellar density is high but not
extraordinary; NGC 5253 for example has an estimated 1 million stars
within a region 1 by 2 pc \citep{TBH00}.  However II~Zw~40 is a tiny galaxy.
\citet{SS70}, \citet{JL} and \citet{Ve96} all
point out that the OB mass of II~Zw~40 is a remarkably high (10\%) fraction
of its total stellar mass, which \citet{Ve96} estimate at $\rm M\sim 2\times
10^7~M_\odot$.  

The 75~pc extent of the starburst core places constraints
on the trigger for the starburst. The crossing time for a disturbance
moving at the sound speed in $10^4$~K gas is roughly 5-10~Myr. While
it is difficult to place ages on these nebulae, comparison with 
Galactic HII regions of similar properties would give the nebulae
ages of  $<1$~Myr \citep{G01}. The ages of these sources are almost
certainly a small fraction of the crossing time of the region, suggesting
that the star formation was triggered nearly simultaneously.
It seems unlikely, therefore, that
``sequential super star cluster formation" is operating in II~Zw~40.

How does the structure seen in II~Zw~40 compare to that in other
supernebula-dominated systems?   II~Zw~40 is like NGC~5253 and He~2-10
in that the infrared and radio fluxes come from obscured super star
clusters. The II~Zw~40 source is more luminous than the NGC~5253 supernebula
in the radio and infrared and contains at least twice the number of O
stars, but they are distributed in multiple sources over a much larger
volume.  The total stellar content of
the II~Zw~40 source is close to that of the main source A in the dwarf
galaxy He~2-10 \citep{KJ99,B01,VJC02}, which it also resembles in size and
radio and infrared flux. He~2-10 has not been observed with high
resolution so the possibility of discrete sub-clusters within the main
sources is open. Why is the starburst in II~Zw~40 concentrated
within 75~pc, that in
He~2-10 200~pc, that in SBS 0335-052 1000~pc \citep{Dale01,I97},
and in NGC~5253 only 2 pc?  That must depend on the
interactions that triggered the formation of the massive star clusters
in the first place and their subsequent evolution.

\section{Conclusions}

Our high-resolution radio and infrared observations of the starburst dwarf galaxy II~Zw~40 
have found that the radio emission at 6, 3.6, and 2~cm is thermal free-free emission with $\sim$70-75\%, of the cm-wave emission confined to a region
less than 3\arcsec\ in diameter.
Within this region, there is a bright core of diameter
$\sim 1.5$\arcsec\ (75 pc) containing at least four sub-sources.
The spectrum of the compact emission appears to
be rising between 3.6 cm and 2 cm, suggesting that the free-free emission
is optically thick at 2~cm, with implied emission measure of 
$\rm EM\sim 10^9~cm^{-6}\,pc$ and electron density
$\rm n_e\sim 3-4\times 10^4~cm^{-3}$. For optically thick
emission, the observed brightnesses
require that the emitting sources are $\sim$1 pc diameter HII regions,
with $\rm N_{Lyc}\sim 6\times 10^{51}~s^{-1}$. 
The 75 pc ``core" starburst 
region  is a complex of four compact ``supernebulae", each  
powered by $\sim$600 O stars, with an estimated $\rm \sim 10^5~M_\odot$ 
in young stars in each cluster. 
The supernebulae/clusters are separated by $\sim$12~pc.  
The total ionization of II~Zw~40 is at least
$\rm N_{Lyc}=1.4\times10^{53}~s^{-1}$, or 14,000 O7 star equivalents. For a Salpeter ZAMS IMF, we estimate that roughly 1-10 million young stars
and a young stellar mass of $\rm 2\times10^6~M_\odot$ are
present within II~Zw~40. The total young stellar luminosity is 
$\rm L_{OB}\sim 3\times 10^9~L_\odot$. The OB luminosity
is in good agreement with or perhaps even higher than the observed infrared 
luminosity of $\rm L_{IR}\sim 1.9\times 10^9~L_\odot$. 
The radio-infrared ratios in this galaxy differ from the usual starburst 
values; we attribute this to
the extreme youth of the star formation activity, the low metal content of the gas, and a very 
weak cool dust component. 
 
\acknowledgments

We are grateful to S. Van Dyk for assistance with the 6~cm data and to
an anonymous reviewer for helpful comments.
This work was supported in part by NSF grant AST-0071276 to J.L.T. and
by the Israel Academy Center for Multi-Wavelength Astronomy grant to S.C.B. 
This research has made use of the NASA/IPAC Extragalactic Database (NED) 
which is operated
by the Jet Propulsion Laboratory, California Institute of Technology, under contract
with the National Aeronautics and Space Administration. The authors wish 
to recognize the significant cultural role that the summit
of Mauna Kea has had within the indigenous Hawaiian community and are
grateful for the opportunity to conduct observations from this mountain.

\clearpage
\figcaption [] {Radio and infrared images of II~Zw~40. Contours are 
$\pm 2^{n/2} \times 0.27$ mJy/beam ($\sim 3\sigma$), starting at n=0. Beams
are plotted in the lower right corners.
{\it a) top, left:} 2~cm VLA map of II~Zw~40. 
Beam is 0\farcs22 x 0\farcs14, p.a. 46$^\circ$.  
{\it b) top, right} 3.6 cm map.
Beam is 0\farcs48 x 0\farcs25, p.a. 52$^\circ$.   
{\it c) bottom, left:} 6 cm map of II Zw 40. 
Beam is 0\farcs89 x 0\farcs43, p.a. 52$^\circ$. 
{\it d) bottom, right:} 6~cm contours atop the $11.7\mu$m LWS image 
of II~Zw~40. The seeing in the IR image is estimated to be 0\farcs2
to 0\farcs5. 
}

\figcaption[]{Radio and infrared spectral energy distributions
for II~Zw~40. {\it a)  left:} Radio SED of II~Zw~40. Crosses with error
bars: single dish measurements of \citet{J78}, \citet{KWT84}, 
\citet{KWB91}, \citet{D93}.
Filled triangles: total fluxes from our naturally-weighted images.
Open triangles: fluxes from the images with (u,v) range restricted
to match the 2~cm (u,v) coverage (see text). Statistical uncertainties are
$\rm \pm 1~mJy~bm^{-1}$; systematic uncertainties due to sidelobes
are estimated to be 50\% larger. The single dish data are fit with an 
optically thin bremsstrahlung spectrum, $\rm S\propto \nu^\alpha,~\alpha
= -0.1$, with contributions of 21 mJy at 6~cm and 12 at 92 cm. 
The nonthermal synchrotron component is fit with $\alpha=-0.75.$ 
{\it b) right:} IR SED of II~Zw~40. Fluxes and references are in
Table~2.}

\figcaption[]{{\it a) top, left:} Optical image of
II~Zw~40 from \citet{BST}. {\it b) top, right:}
The $11.7\mu$m LWS image of II~Zw~40. The entire frame is about 10\arcsec\
across. {\it c) bottom, left:} Blue optical image of II~Zw~40 \citep{BST}.
{\it d) bottom, right:} LWS image placed atop the blue optical image to the 
same scale. The LWS inset is $\sim$10\arcsec\ square.}

\figcaption [] {{\it a) left:} Uniformly-weighted
2~cm VLA map of II~Zw~40, in greyscale, with naturally-weighted
2~cm map (see Figure 1) overlaid in contours.  Greyscale
range is 0.4-0.6 mJy, $\sim$2.5-4$\sigma$. Uniformly-weighted beam 
is 0\farcs14 x 0\farcs12, p.a. 45$^\circ$. Contours are $2^{n/2}$
times $\pm$0.4 mJy/beam ($\sim 6\sigma$ for the naturally-weighted map), 
starting at n=0. 
{\it b) right:} Uniformly-weighted 2~cm map. Contours as for a),
$\sim 2.5\sigma$ for this map.}

\clearpage
\begin{deluxetable}{lcccccccc}
\rotate
\tabletypesize{\scriptsize}
\tablenum{1}
\tablewidth{0pt}
\tablecaption{II~Zw~40 Radio Fluxes}
\tablehead{
\colhead{$\lambda$} 
&\colhead{Single Dish}
&\colhead{r.m.s.}
&\colhead{Beam} 
&\colhead{$\theta_{max}^b$}
&\colhead{Peak Flux}
&\colhead{Total Mapped}
&\colhead{Fraction of}
&\colhead{``Matched" Beam$^d$}
\nl
\colhead{} 
&\colhead{Flux$\rm ^a$}
&\colhead{}
&\colhead{} 
&\colhead{}
&\colhead{}
&\colhead{Flux$\rm ^{b,c}$}
&\colhead{Single Dish Flux}
&\colhead{Fluxes}
\nl
\colhead{} 
&\colhead{(mJy)}
&\colhead{(mJy/bm)}
&\colhead{\arcsec$\times$\arcsec, pa$^\circ$}
&\colhead{(FWHM)}
&\colhead{(mJy/beam)}
&\colhead{(mJy)} 
&\colhead{(mJy)}
&\colhead{(mJy)}
\nl
}
\startdata
6~cm &22&0.11&0.89 x 0.43, 52& 10\arcsec&3.4 
&15$\pm$1 &0.68&9$\pm$1.5\nl
3.6~cm&21&0.088&0.48 x 0.25, 52&7\arcsec& 1.4
&12$\pm$1&0.57&10$\pm$1.5\nl
2~cm &18.5 &0.10&0.21 x 0.14, 45 &4\arcsec & 0.64
&14$\pm$ 1.5&0.75&14$\pm$1.5\nl

\enddata

\tablenotetext{a}{Extrapolated from values at 4.9, 10.7, and 
24.5 GHz. \citet{KWB91}, \citet{KWT84}, \citet{J78}}.

\tablenotetext{b}{$\theta_{max}$ is the maximum sizescale that is
well sampled by these images. Fluxes and peak fluxes are therefore
lower limits to the total flux.}

\tablenotetext{c}{Total fluxes are obtained by integrating over
a circular aperture centered on the peak flux in naturally-weighted
maps. }

\tablenotetext{d}{``Matched" beam fluxes were obtained by making 
maps with (u,v) data restricted to baselines $>$ 20k$\lambda$; these
maps are sensitive to structures smaller than 4\arcsec\ only. See text. }

\end{deluxetable}

\clearpage
\begin{deluxetable}{lcccccccc}
\tablenum{2}
\tablewidth{0pt}
\tablecaption{Infrared Continuum Spectrum}
\tablehead{
\colhead{} 
&\colhead{10.1 $\mu$m}
&\colhead{10.4 $\mu$m}
&\colhead{11.7 $\mu$m} 
&\colhead{12 $\mu$m}
&\colhead{20 $\mu$m} 
&\colhead{25 $\mu$m}
&\colhead{60 $\mu$m}
&\colhead{100 $\mu$m}  
\nl
}
\startdata
Flux (Jy)&0.20$^a$&0.22$^b$&0.24$^c$&0.46$^d$&1.0$^e$
&1.91$^d$&6.61$^d$&$$5.8$^d$\nl
Uncertainty (Jy)&$0.02$&0.04&0.04&0.05&0.2&0.2&0.7&0.9\nl

\enddata

\tablenotetext{a}{\citet{WWB}}
\tablenotetext{b}{\citet{RL72}}
\tablenotetext{c}{This paper.}
\tablenotetext{d}{\citet{V93}. 
100$\mu$m flux is affected by Galactic cirrus emission.}
\tablenotetext{e}{\citet{R91}}

\end{deluxetable}


\begin{thebibliography}{}
\bibitem[Baldwin, Spinrad, \& Terlevich (1982)]{BST}
Baldwin, J. A., Spinrad, H., \& Terlevich, R. 1982, \mnras, 198, 535
\bibitem[Beck et al.(2001)]{B01} Beck, S. C., Turner, J. L., Ho, Gorjian, 
V. 2001, \aj, 000, 000
\bibitem[Beck et al. (2002)]{B02} Beck, S. C., Crowther, P.,\&Conti, P. 2002, in preparation.
\bibitem[Beck, Turner, \& Gorjian (2001)]{B01}
Beck, S. C., Turner, J. L., \& Gorjian, V. 2001, \aj, 122, 1365
\bibitem[Beck et al.(1986)]{B86} Beck, S. C., Turner, J. L., Ho, P. T. P.
1986, \apj, 309, 70
\bibitem[Beck et al.(1996)]{B96} Beck, S. C., Turner, J. L., Ho, P. T. P., 
Lacy, J. H., \& Kelly, D. 1996, \apj, 457, 610
\bibitem[Brinks \& Klein (1988)]{BK88} Brinks, E., \& Klein, U. 1988,
\mnras, 231, 63P
\bibitem[Calzetti et al.(1997)]{C97} Calzetti, D., Meurer, G. R., 
Bohlin, R. C., Garnett, D. R., Kinney, A. L., 
Leitherer, C., \& Storchi-Bergmann, T. 1997, \aj, 114, 1834
\bibitem[Crowther et al.(1999)]{Cr99}
Crowther, P. A., Beck, S. C., Willis, A. J., Conti, P. S., Morris, P. W., \& 
Sutherland, R. S. 1999, \mnras, 304, 654
\bibitem[Dale et al. (2001)]{Dale01} Dale, D. A., Helou, G., Neugebauer, G.,
Soifer, B. T., Frayer, D. T., \& Condon, J. J. 2001, \aj, 122, 1736
\bibitem[Davies, Sugai, \& Ward (1998)]{DSW} 
Davies, R. I., Sugai, H., \& Ward, M. J. 1998, \mnras, 295, 43
\bibitem[Deeg et al.(1993)]{D93} 
Deeg, H.-J., Brinks, E., Duric, N., Klein, U.,
\& Skillman, E. 1993, \apj, 410, 626
\bibitem[Genzel et al.(1982)]{Genzel82} Genzel, R., Becklin, E. E., 
Wynn-Williams, C. G., Moran, J. M., Reid, M. J., Jaffe, D. T.,
\& Downes, D. 1982, \apj, 255, 527
\bibitem[Gorjian, Turner, \& Beck (2001)]{G01}
Gorjian, V., Turner, J. L., \& Beck, S. C. 2001, \apj, 554, L29
\bibitem[Ho, Beck, \& Turner (1990)]{H90}Ho, P. T. P., Beck, S. C., \&
Turner, J. L. 1990, \apj, 349, 57
\bibitem[Hummer \& Storey (1987)]{HS87}
Hummer, D. G., \& Storey, P. J. 1987, \mnras, 224, 801
\bibitem[Izotov et al. (1997)]{I97}
Izotov, Y. I., Lipovetsky, V. A., Chaffee, F. H., Foltz, C. B.,
Guseva, N. G., \& Knaiazev, A. Y. 1997, \apj, 476, 698
\bibitem[Jaffe, Perola, \& Tarenghi (1978)]{J78}
Jaffe, W. J., Perola, G. C., \& Tarenghi, M. 1978, \apj, 224, 808
\bibitem[Jones \& Puetter (1993)]{JP93}
Jones, B., \& Puetter, R. 1993, Proc. SPIE, 1946, 610
\bibitem[Joy \& Lester (1988)]{JL}Joy, M., \& Lester, D. F. 1988, \apj,
331, 145
\bibitem[Kawara, Nishida, \& Phillips(1989)]{K89}
Kawara, K., Nishida, M., \& Phillips, M. M. 1989, \apj, 337, 320
\bibitem[Klein, Weiland, \& Brinks (1991)]{KWB91} Klein, U., Weiland, H., 
\& Brinks, E. 1991, \aap, 246, 323
\bibitem[Klein, Wielebinski, \& Thuan (1984)]{KWT84}Klein, U.,
Wielebinski, R., \& Thuan, T. X. 1984, \aap, 141, 241
\bibitem[Kobulnicky \& Johnson (1999)]{KJ99} Kobulnicky, H. A., \&
Johnson, K. E. \apj, 527, 154 
\bibitem[Massey \& Hunter(1998)]{MH98}
Massey, P., \& Hunter, D. A. 1998, \apj, 493, 180
\bibitem[Meier, Turner, \& Beck (2002)]{M02}Meier, D. S., Turner, J. L.,
\& Beck, S. C. 2002, \aj, in press
\bibitem[Moorwood \& Oliva (1988)]{MO}
Moorwood, A. F. M., \& Oliva, E. 1988, \aap, 203, 278
\bibitem[Rieke \& Lebofsky (1985)]{RL85}
Rieke, G. H., \& Lebofsky, M. J. 1985, \apj, 288, 618
\bibitem[Rieke \& Low (1972)]{RL72} 
Rieke, G., H. \& Low, F. J. \apj, 176, L95
\bibitem[Roche et al. (1991)]{R91}
Roche, P. R. Aitken, D. K., Smith, C. H., \& Ward, M. J. 1991, \mnras, 
248, 606
\bibitem[Sage et al. (1992)]{S92} Sage, L. J., Salzer, J. J., Loose,
H.-H., \& Henkel, C. 1992, \aap, 265, 19
\bibitem[Sargent (1970)]{S70} 
Sargent, W. L. W. 1970, \apj, 160, 405
\bibitem[Sargent \& Searle (1970)]{SS70} 
Sargent, W. L. W., \& Searle, L. 1970, \apj, 162, 151
\bibitem[Sramek \& Weedman (1986)]{SW}
Sramek, R. A., \& Weedman, D. W. 1986, \apj, 302, 640
\bibitem[Tarchi et al. (2000)]{Tar00}
Tarchi, A., Neininger, N., Klein, U., Greve, A., Garrington, S. T.,
Muxlow, T. W. B., Pedlar, A., \& Glendenning, B. E. 2000, \aap, 358, 95
\bibitem[Thornley et al. (2000)]{Thornley}
Thornley, M., Schreiber, N. M. F., Lutz, D., Genzel, R., Spoon, H. W. W.,
Kunze, D., \& Sternberg, A. 2000, \apj, 539, 641
\bibitem[Tremonti et al. (2001)]{Tre01}
Tremonti, C. A., Calzetti, D., Leitherer, C., \& Heckman, T. M. 2001,
\apj, 555, 322
\bibitem[Turner, Ho, \& Beck (1998)]{THB98}
Turner, J. L., Ho, P. T. P., \& Beck, S. C. 1998, \aj, 116, 1212
\bibitem[Turner, Beck, \& Ho (2000)]{TBH00}
Turner, J. L., Beck, S. C., \& Ho, P. T. P. 2000, \apj, 532, L109
\bibitem[Vacca \& Conti (1992)]{VC} Vacca, W. D., \& Conti, P. S. 1992,
\apj, 401, 543
\bibitem[Vacca, Johnson, \& Conti (2002)]{VJC02}
Vacca, W. D., Johnson, K. E., \& Conti, P. S. 2002, \aj, 123, 772
\bibitem[Vader et al. (1993)]{V93} 
Vader, J. P., Frogel, J. A., Terndrup, D. M., \& Heisler, C. A.
1993, \aj, 106, 1743
\bibitem[van Zee, Skillman, \& Salzer (1998)]{Van98} van Zee, L., Skillman, 
E. D., \& Salzer, J. 1998, \apj, 116, 1186
\bibitem[Vanzi et al. (1996)]{Ve96}
Vanzi, L., Rieke, G. H., Martin, C. L., \& Shields, J. C. 
1996, \apj, 466, 150
\bibitem[Walsh \& Roy(1993)]{WR}
Walsh, J. R., \& Roy, J.-R. 1993, \mnras, 262, 27
\bibitem[Wynn-Williams \& Becklin (1986)]{WWB}
Wynn-Williams, C. G., \& Becklin, E. E. 1986, \apj, 308, 620

\end{thebibliography}
\end{document}